# Water ice at low to midlatitudes on Mars

Mathieu Vincendon[1], François Forget[2], and John Mustard[1]

[1]*Department of Geological Sciences, Brown University, Providence, RI, USA.* [2]*Laboratoire de Météorologie Dynamique, Université Paris 6, Paris, France.*

**Abstract: In this paper, we analyze water ice occurrences at the surface of Mars using near-infrared observations, and we study their distribution with a climate model. Latitudes between 45°S and 50°N are considered. Data from the *Observatoire pour la Minéralogie, l'Eau, les Glaces et l'Actitité* and the *Compact Reconnaissance Imaging Spectrometer for Mars* are used to assess the presence of surface water ice as a function of location and season. A modeling approach combining the 1-D and 3-D versions of the General Circulation Model of the *Laboratoire de Météorologie Dynamique de Jussieu* is developed and successfully compared to observations. Ice deposits 2-200 μm thick are observed during the day on pole facing slopes in local fall, winter and early spring. Ice extends down to 13° latitude in the Southern Hemisphere but is restricted to latitudes higher than 32° in the north. On a given slope, the pattern of ice observations at the surface is mainly controlled by the global variability of atmospheric water (precipitation and vapor), with local ground properties playing a lower role. Only seasonal surface ice is observed: no exposed patches of perennial ground ice have been detected. Surface seasonal ice is however sensitive to subsurface properties: the results presented in this study are consistent with the recent discovery of low latitude subsurface ice obtained through the analysis of $CO_2$ frost.**

1. Introduction

Water on Mars follows a complex seasonal cycle that had varied over geologic times. While liquid water has been involve in the early processing of the martian surface [*Bibring et al.*, 2006; *Carr*, 1979], it is not observed today, although it could be locally stable [*Haberle et al.*, 2001; *Hecht*, 2002; *Zorzano et al.*, 2009]. So far, water on Mars has only been detected in the form of vapor, ice, adsorbed molecules, or as a component of minerals [*Bibring et al.*, 2005; *Farrell et al.*, 2009; *Jouglet et al.*, 2007; *Kieffer et al.*, 1976; *Spinrad and Richardson*, 1963]. The most prominent reservoirs of martian water are ice, mainly located in the northern and southern polar deposits and in the high-latitude permafrost [*Clifford et al.*, 2000; *Feldman et al.*, 2004]. For comparison the reservoir of water vapor is about 6 orders of magnitude lower [*Montmessin et al.*, 2004].

The low to midlatitude regions of Mars are a key environment to study the distribution of surface water ice and its role in the processing of the Martian surface. The stability limits of both seasonal ice and perennial near-surface ice is observed at those latitudes [*Carrozzo et al.*, 2009; *Feldman et al.*, 2004; *Schorghofer and Edgett*, 2006; *Vincendon et al.*, 2010]. As attested by numerous morphological features [*Head et al.*, 2003; *Mustard et al.*, 2001; *Squyres and Carr*, 1986] and climate modeling predictions [*Forget et al.*, 2006; *Levrard et al.*, 2004; *Mischna et al.*, 2003; *Schorghofer*, 2007], these limits has undergone strong variations over geological epochs. Colder, "ice age" era have seen the formation of several landforms still observed today [*Head et al.*, 2003; *Mustard et al.*, 2001; *Neukum et al.*, 2004; *Schon et al.*, 2009], and mineralogical assemblages found at those latitudes could have been influenced by water ice [*Mangold et al.*, 2010; *Niles and Michalski*, 2009]. This low to midlatitudes water ice, with its potential for melting when the obliquity varies [*Christensen*, 2003; *Costard et al.*, 2002; *Williams et al.*, 2009], or at present day [*Kreslavsky and Head*, 2009; *Mohlmann*, 2008], has been hypothesized to be a conceivable habitable environment on Mars [*Morozova et al.*, 2007; *Pavlov et al.*, 2009; *Price*, 2007].

Studying surface or subsurface water ice at low to midlatitudes from remotely sensed data is subject to some limitations and constraints. The simultaneous presence of water ice clouds and surface $CO_2$ frost complicates the detectability of surface water ice. Spatial resolution is also a limitation, as favorable conditions required for surface or subsurface ice to be stable may be of limited spatial extent. The imaging spectrometers OMEGA (*Observatoire pour la Minéralogie, l'Eau, les Glaces, et l'Activité*) and CRISM (*Compact Reconnaissance Imaging Spectrometer for Mars*) are well suited for the study of water ice as they offer $CO_2$/$H_2O$ and cloud/frost separation capabilities, combined with high spatial resolution (down to 350 meters per pixel for OMEGA and 18 meters per pixel for CRISM). The first Mars year of OMEGA observation have been successfully used to study water frost deposits in the southern hemisphere equatorward of 30° [*Carrozzo et al.*, 2009], and subsurface water ice in the 45°S – 25°S latitude range has been discovered through the analysis of OMEGA and CRISM detections of $CO_2$ frost [*Vincendon et al.*, 2010].

In this paper, we perform the mapping of surface water ice from the edge of the polar caps (about 45°-50° latitude) to the equator, in both hemispheres, using OMEGA and CRISM data covering more than 3 Mars years (section 2). We then develop a new modeling approach based on the LMD (*Laboratoire de Météorologie Dynamique*) GCM (*General Circulation Model*) used to predict formation of ice at these latitudes (section 3). Observations and modeling results are then compared and discussed in section 4.

## 2. Observations

The OMEGA and CRISM instruments are imaging spectrometers observing in the visible and near-infrared wavelengths with a spectral sampling between 7 nm and 40 nm. Since early 2004, OMEGA has acquired near-global coverage with an average resolution of 1 km per pixel. Because of the orbit of Mars Express, the actual resolution varies between 5 and 0.35 km per pixel and many regions have been repeatedly observed. Since late 2006, CRISM has acquired observations of Mars with a spatial resolution as high as 18 m per pixel in the targeted mode, covering a few percentages of the planet, and 230 m per pixel in the low resolution mode (which plans to acquire global coverage).

## 2.1. Data reduction

These observations are sensitive to the presence of a few μm of water ice within the path of electromagnetic radiation reflected from the surface of Mars [*Langevin et al.*, 2007]. Water ice notably absorbs photons around 1.5 and 2 μm, creating characteristic broad absorption features several tenths of μm wide. In the low latitudes, these spectral features are typically a few percentages deep [*Carrozzo et al.*, 2009]. Algorithms to conduct a global, automated search for the presence of these features can be easily implemented. Such an approach, however, highlights several false detections in addition to real surface deposits we are looking for. Among the caveats are the complex structure of the random spectral noise, data anomalies, wavelength-dependent nonlinearity in detectors response, water ice clouds, and surface deposits of minerals such as sulfate, which absorb radiation at similar wavelengths. As a consequence, automatic processing provides a first analysis of the global data and positive detections must be validated in detail. Spectral analysis is used to confirm the presence of water ice, while analyzing the geologic setting makes it possible to distinguish between surface and atmospheric origins.

It has been possible to reach a high level of confidence in the automatic detection algorithm applied to the OMEGA dataset, as detailed in Figure 1. Only a few false positive detections that must be manually removed remain after applying this algorithm. They are due to surface minerals with unusual deep spectral features, such as the olivine spots seen in Nili Fossae, or to small extent topographic clouds such as those seen in Peta crater. Due to the much more complex structure of the CRISM noise [*Parente*, 2008], a similar approach has not been implemented for this data set: all positive CRISM detections have been checked manually.

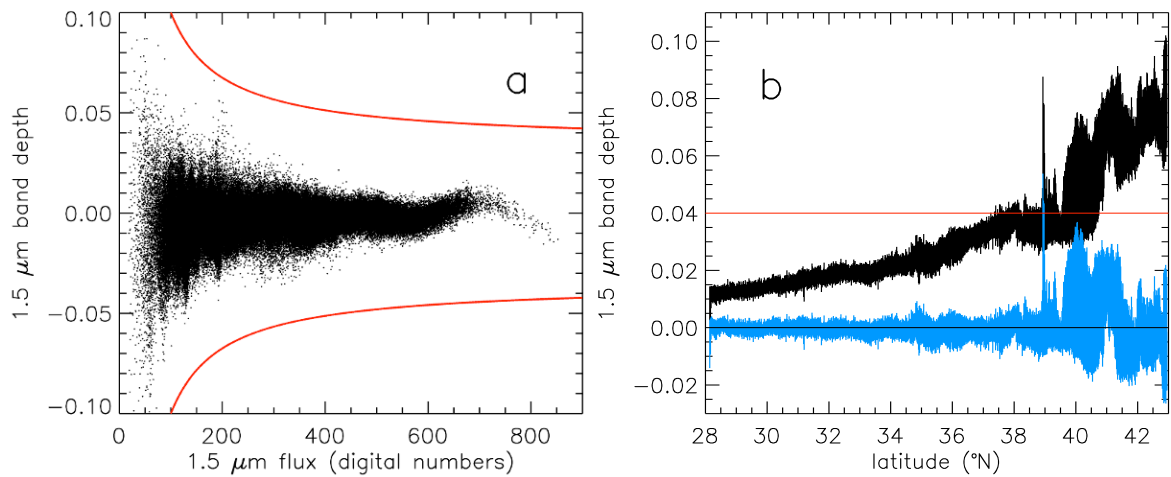

**Figure 1.** *OMEGA detection algorithm. The 1.5 μm water ice band depth is estimated using the criteria, $depth = 1 - R(1.50\mu m)/\left(R(1.39\mu m)^{0.7} \times R(1.77\mu m)^{0.3}\right)$. The noise in OMEGA data is dominated by the read noise (1.85 DN). The noise limit is therefore of the form, $depth > (C \times 1.85)/flux$, with $C$ a constant. To account for non-linearity issues around 1.5*

*µm, an offset is added. In panel a, the selected detection limit (depth > 6.5/ flux + 0.035, about 4% or higher) has been empirically determined based on observations without ice (dots). Isolated spikes due to cosmic rays or data anomalies are removed by selecting only detections of more than two contiguous pixels. Extreme photometric conditions (i >80°, e >40°) are not considered due to aerosols scattering. As ice deposits are observed on pole facing slopes, most ice clouds of moderate extent are removed by considering only detections with a spatial extent smaller than 5 km in latitude and 50 km in longitude. Finally, the background atmospheric ice opacity due to large ice clouds is removed by dividing the observed 1.5 µm depth by the latitudinal trends in 1.5 µm depth. An example is shown in Figure 1b: the observed band depth (black) increases with latitude due to increasing cloud opacity. The corrected band depth (lighter tone/blue, centered on 0) is above the noise limit (4%) only at 39°N due to surface ice localized on a pole facing slope.*

### 2.2. Spatial and seasonal distribution

Surface water ice deposits are mapped in Figure 2. In the Southern Hemisphere, the distribution of ice strongly varies with longitude. Ice is detected down to 13°S around 10°E and on the northern wall of Valles Marineris near 60°W, while it is not observed equatorward of 40°S at 20°W. The major concentrations of surface ice deposits are south of the great volcanoes and west of the Hellas basin. A particularly dry area is observed between 50°W and 0°. In the Northern Hemisphere, the distribution of ice is more homogeneous, but ice is not observed equatorward of 32°N.

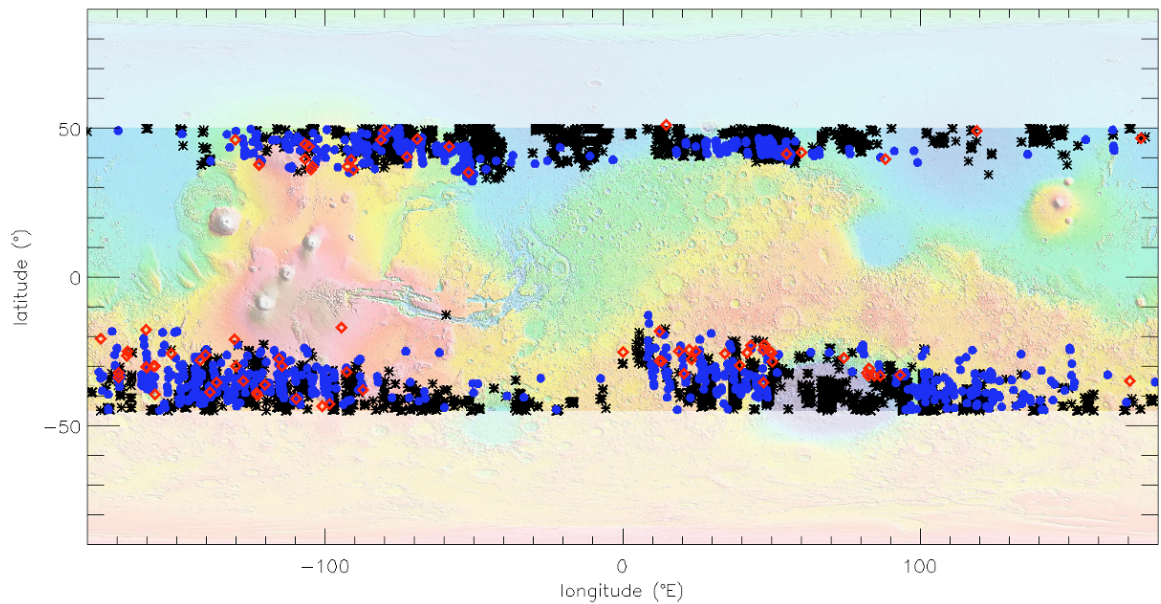

**Figure 2.** *Observed surface water ice deposits detected between 45°S and 50°N shown on a MOLA shaded relief map. OMEGA, black stars; CRISM Low-Res ("MRDR"), blue dots; CRISM High-Res ("FRT", "HRS" and "HRL"), red diamonds. Ice is observed down to 13°S and 32°N.*

The latitude-season diagrams of ice deposits are shown in Figure 3. Ice is observed only during local fall and winter (and early spring for the Northern Hemisphere). These deposits correspond well to expected seasonal frost. We see no deposits that can be interpreted as perennial ice. This is consistent with the expected rapid sublimation of snowpack at those latitudes [*Byrne et al.*, 2009; *Williams et al.*, 2008]. The distribution of frost derived from observations in the southern hemisphere generally agrees with the results of [*Carrozzo et al.*, 2009] when they overlap in time and location (i.e., for $L_S$ in the 100°–150° range and for latitudes in the 15°S–30°S range), except for their detections at $L_S$ 20°S and 30°S which are due to clouds according to our study. These authors have observed water ice detection in the northern hemisphere at latitudes lower than 30°N in summer, which where not considered reliable due to the potential contribution of clouds. These detections are not present in our results, which confirm their atmospheric origin.

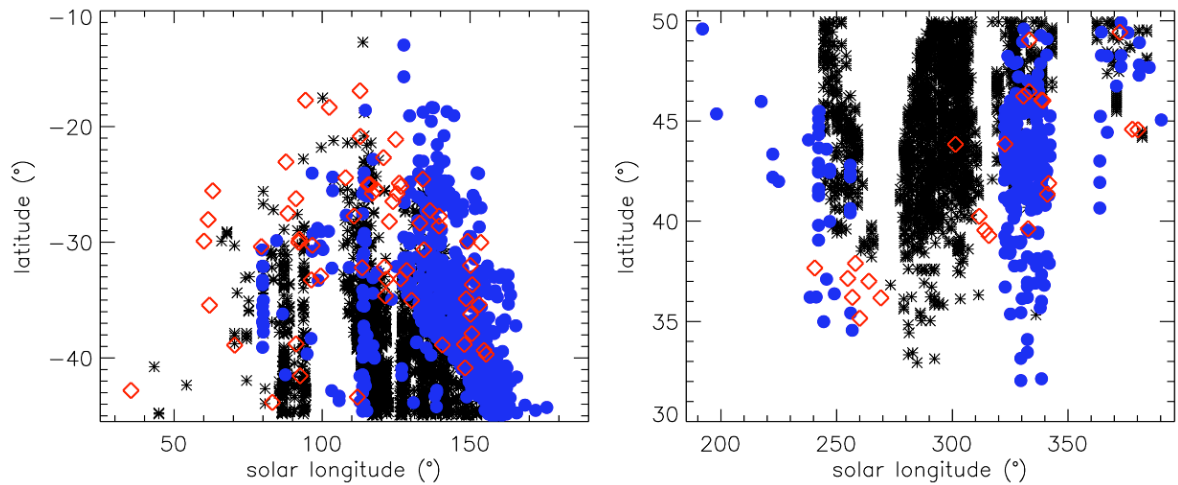

**Figure 3.** *Observed surface water ice deposits detected between 45°S and 50°N shown in a latitude/season diagram (left, south; right, north). OMEGA, black stars; CRISM L-R, blue dots; CRISM H-R, red diamonds. At those latitudes water ice is observed only during local autumn and winter (as well as during early spring for the Northern Hemisphere).*

### 2.3. Observational bias

Since early 2004, almost all the midlatitudes have been observed at appropriate winter $L_S$ by either CRISM or OMEGA, which rule out any major observational bias. Spatial resolution is not a major issue: the spatial and temporal stability pattern is similar at first order as seen in the three datasets (OMEGA, with its resolution in the 0.35–5 km range; CRISM LR, at 230 m per pixel resolution and CRISM HR, with a 18 m per pixel resolution; see Figures 2 and 3). Second order, slight differences are observed: at the most equatorward latitudes, a greater number of deposits are observed that result from the higher spatial resolution of CRISM (Figure 2). As a result of the orbital properties of Mars Express, southern midlatitudes have not been observed by OMEGA at $L_S > 130–150°$. A similar bias is observed prior to $L_S$ 240° in the Northern Hemisphere.

Local time is not a major issue in the distribution of frost presented in this study. All CRISM observations are acquired around 3 pm. OMEGA observation can be obtained anytime in daylight. However, we do not consider early morning and late evening observations as we restrict ourselves to solar zenith angles lower than 80° (above that limit the contribution of light scattered by aerosols dominates, see e.g. [*Vincendon et al.*, 2007]). No specific variability has been found in the OMEGA data set for hours surrounding midday.

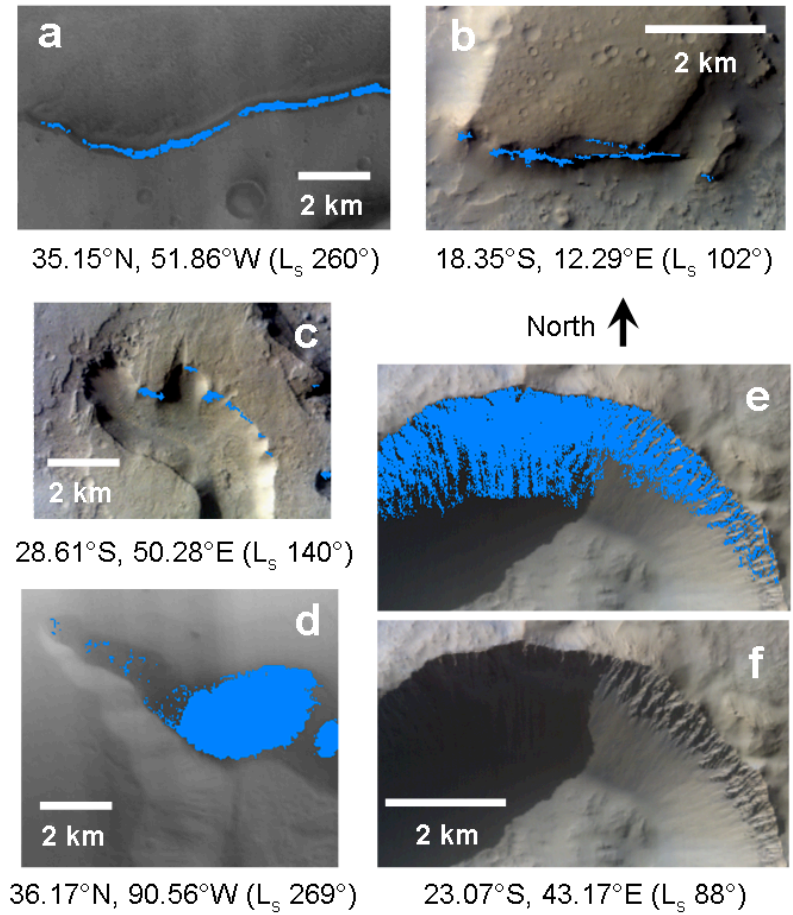

**Figure 4.** *CRISM high-resolution observations (18-40 m per pixels) of low to midlatitude water ice deposits. (a-e) Ice detected using the 1.5 μm water ice feature is mapped in blue above a background visible composite image at five locations. (f) Observation "e" is also shown without ice to highlight the geologic setting. Ice is observed during local winter in both the (a, d) Northern and (b, c, e) Southern hemispheres. Ice forms on pole facing slopes linked with all kind of geomorphology: (a) escarpments, (b, d) mesas, (c) valleys, and (e) crater rims.*

### 2.4. Local setting

Deposits are found on steep slopes facing the pole of their local hemisphere (Figure 4). The coverage of crater rims by frost can extend to east and west facing walls when small-scaled topographic features such as gullies creates local pole-facing slopes (Figure 4e; see also [*Dickson and Head*, 2009]). In general, frost is observed on any kind of pole facing slopes (crater walls, mesas, knobs, etc.), regardless of the underlying material. Slopes on which water ice forms are typically in the 20°–30° range according to MOLA topography measurements [*Kreslavsky and Head*, 2000; *Smith et al.*, 1999; *Vincendon et al.*, 2010]. At the edge of the stability pattern (lowest latitudes, earliest/latest $L_S$), frost tends to be observed mainly on the steepest slopes (30°).

### 2.5. Link with gullies

Several mechanisms could be responsible or participate to the formation and modification of Martian gullies (see, e.g., a review by [*Dickson and Head*, 2009]), including melting of surface water ice during period of high obliquity or at present day [*Dickson and Head*, 2009; *Hecht*, 2002; *Williams et al.*, 2009]. Variations in shape and color have been observed in gully channels during the last years [*Dundas et al.*, 2010; *Malin et al.*, 2006], which correspond to the years of operation of OMEGA and CRISM. In this study, we found frost to be present at the location of several gully clusters as mapped by [*Dickson et al.*, 2007; *Heldmann and Mellon*, 2004]. The distribution of gullies does, however, not necessarily correlate with the distribution of present-day surface frost. For example, gullies are frequently observed in the dry region at the east of Argyre (30°W–0° / 45°S–30°), while not found in the ice-rich 20°–30° latitude band. Interestingly, the location of recent gully activity determined by [*Dundas et al.*, 2010] corresponds to areas of water frost stability. $CO_2$ frost is also present at most of these sites of recent gully activity (see [*Vincendon et al.*, 2010]).

## 3. Modeling

### 3.1. Modeling approach

Our modeling approach is based on the General Circulation Model (GCM) of the *Laboratoire de Météorologie Dynamique de Jussieu* (LMD) [*Forget et al.*, 1999]. This GCM has been extensively validated through comparisons with observations, notably observations related to the water cycle [*Forget et al.*, 2008; *Montmessin et al.*, 2004]. The main version of this GCM is a 3-D grid point dynamic model, from which a 1-D version has been derived. The same physics is used in this 1-D version. In 1-D, however, it is possible to explore the environment on any slope with any orientation. Local properties, including those varying with season such as the aerosols optical depth, have to be prescribed. This 1-D model is therefore suited for the study of local phenomena that cannot be simulated by the global low-resolution model and for which feedbacks on the large-scale properties of the meteorology can be neglected. An illumination scheme designed for surface slopes has been recently included in this 1-D code [*Spiga and Forget*, 2008] and successfully used to analyze observations of $CO_2$ ice [*Vincendon et al.*, 2010]. Modeling the condensation of ice on Mars is trickier for $H_2O$ than for $CO_2$. $CO_2$ is the main component of Mars atmosphere (95%): the partial pressure of $CO_2$ is roughly equal to the total pressure. $CO_2$ condenses on the ground when the temperature of the surface reaches the frost point, a pressure-dependent parameter. Seasonal and spatial variations of pressure can be simply considered: they are primarily due to global change in atmospheric mass resulting from the seasonal condensation/sublimation of $CO_2$ in the polar regions, modulated by the topography [*Hourdin et al.*, 1995]. The situation is more complex for $H_2O$, a minor component of the Martian atmosphere (<0.1%). At Martian pressure, water can be found in the ice phase when surface temperatures are below 260K – 280K, which is common on Mars. The situation is then similar to liquid water on Earth: two mechanisms can lead to the formation of water ice on the ground. First, direct deposition of water vapor from saturated air (i.e., formation of "frost") occurs when the partial vapor pressure is higher than the saturation vapor pressure at surface temperature. The second mechanism is precipitation of water ice particles that have condensed higher in the atmosphere. Ice formed this way will be stable if the partial pressure of vapor is higher than the saturation vapor pressure; otherwise it will progressively sublimate with a flux proportional to the wind speed and to the difference between the partial and saturation vapor pressures [*Madeleine et al.*, 2009; *Montmessin et al.*, 2004]. As a consequence, modeling water ice requires a good knowledge of the partial vapor pressure and of the amount of precipitation, parameters which are driven by a complex meteorological dynamic on Mars [*Montmessin et al.*, 2004]. We have constructed global maps of the partial vapor pressure and the amount of precipitation with a time step of half an hour using the 3-D GCM. The spatial sampling of these maps is 300 km. Values from these lookup tables are then used as inputs to the 1-D model, which locally computes the temperature on slopes at a given location as a function of time. The main assumption behind this approach is that the presence of frost patches on slopes does not significantly impact the amount of water vapor or water precipitation in the boundary layer. This hypothesis is relevant considering the small extent of these isolated deposits (deposits a few kilometers wide separated by distances greater by more than one order of magnitude) compared to typical atmospheric movements (winds of several meters per second).

### 3.2. Selected default parameters

In addition to the water vapor pressure and the ice sedimentation flux, two other parameters that vary with latitude, longitude, and season are first computed using the 3-D GCM: the total pressure (so as to account for second order mechanisms affecting total pressure [*Hourdin et al.*, 1993]) and the temperature of flat surfaces surrounding slopes.

The ground (below potential ice deposits) is parameterized by its surface albedo and by two layers of different thermal inertia. The depth of the boundary between the two layers can be selected. When not otherwise specified, the two layers have the same thermal inertia, and the albedo and thermal inertia as a function of latitude and longitude correspond to average trends derived from TES retrievals [*Putzig et al.*, 2005] ; a subsurface layer with the thermal inertia of water ice (2120 kg $K^{-1}$ $s^{-5/2}$) is used in the Southern Hemisphere at latitudes higher than 25°, using the latitude dependent depth of [*Vincendon et al.*, 2010].

Longitudes 10°E in the Southern Hemisphere and 50°W in the northern hemisphere are considered in more detail in section 4: the default surface inertia and surface albedo are 250 kg K$^{-1}$ s$^{-5/2}$ and 0.15, respectively, for these longitudes. An angle of 30° and a pole facing orientation are considered for slopes geometric properties.

Two parameters describe ice properties: the solar albedo and the infrared emissivity. The water ice/frost emissivity is between 0.95 and 1.00 across the thermal infrared region [*MODIS UCBS Emissivity library*]: a value of 1.00 will be used as results are weakly affected over that range. The solar albedo of water ice is known to vary significantly on Mars, from low values of about 0.25 to higher values of 0.6 depending on dust contamination and grain size [*Langevin et al.*, 2005]. Assessing the albedo of ice deposits on slopes is especially difficult due to the complex lighting and viewing geometries. A standard value of 0.4 [*Montmessin et al.*, 2004] is a good proxy for winter water frost observed at higher latitudes on flat surfaces [*Vincendon et al.*, 2007]. The model also accounts for the condensation and sublimation of $CO_2$ ice: an albedo of 0.65 and an emissivity of 1 are selected [*Vincendon et al.*, 2010].

The optical depth of aerosols corresponds to the annual variations seen by Opportunity during MY27, scaled by surface pressure ("well-mixed conditions" hypothesis). These variations were shown to provide a relevant estimate of the optical depth at all longitudes for low to midlatitudes [*Vincendon et al.*, 2009]. As indicated previously, the sublimation flux of water is proportional to the wind speed: a value of 20 m s$^{-1}$ is selected, as mesoscale modeling indicates winds in the 10 – 30 m s$^{-1}$ range on slopes [*Spiga et al.*, 2010].

Below we present model results corresponding to the equatorward limit of the presence of water ice at a local time of 3 PM as a function of $L_S$. We used in section 2.1 a detection limit of about 4% for the 1.5 µm band depth. Considering laboratory measurements of near-IR spectra of ices [*Schmitt et al.*, 2004] and the decreasing spectral contrast of surface features when seen through the aerosols layer [*Vincendon et al.*, 2007], we should be sensitive to deposits a few µm thick. The model predicts the mass of ice deposits, from which the thickness can be estimated assuming a water ice density of 920 kg m$^{-3}$. When not otherwise specified, a 2 µm thickness threshold will be considered in the model.

We analyze in the next section the impact of changing assumptions regarding most of these parameters: water ice albedo, aerosols optical depth, surface/subsurface thermal inertia, depth of the subsurface layer, surface albedo, slope angle, wind, and thickness threshold.

## 4. Results

Observations and modeling predictions are mapped at different $L_S$ in Figure 5 and 6. Overall, the stability pattern predicted by the model using the default set of parameters is consistent with most observations for both the Southern and the Northern hemisphere. We first analyze the global variability of ice stability in subsections 4.1 and 4.2. Afterwards, we study local slope properties using latitude-season diagrams (subsections 4.3 to 4.5). In these sections longitudes corresponding to the equatorward limit of water ice are selected: 10°E for the Southern Hemisphere and 50°W for the Northern Hemisphere.

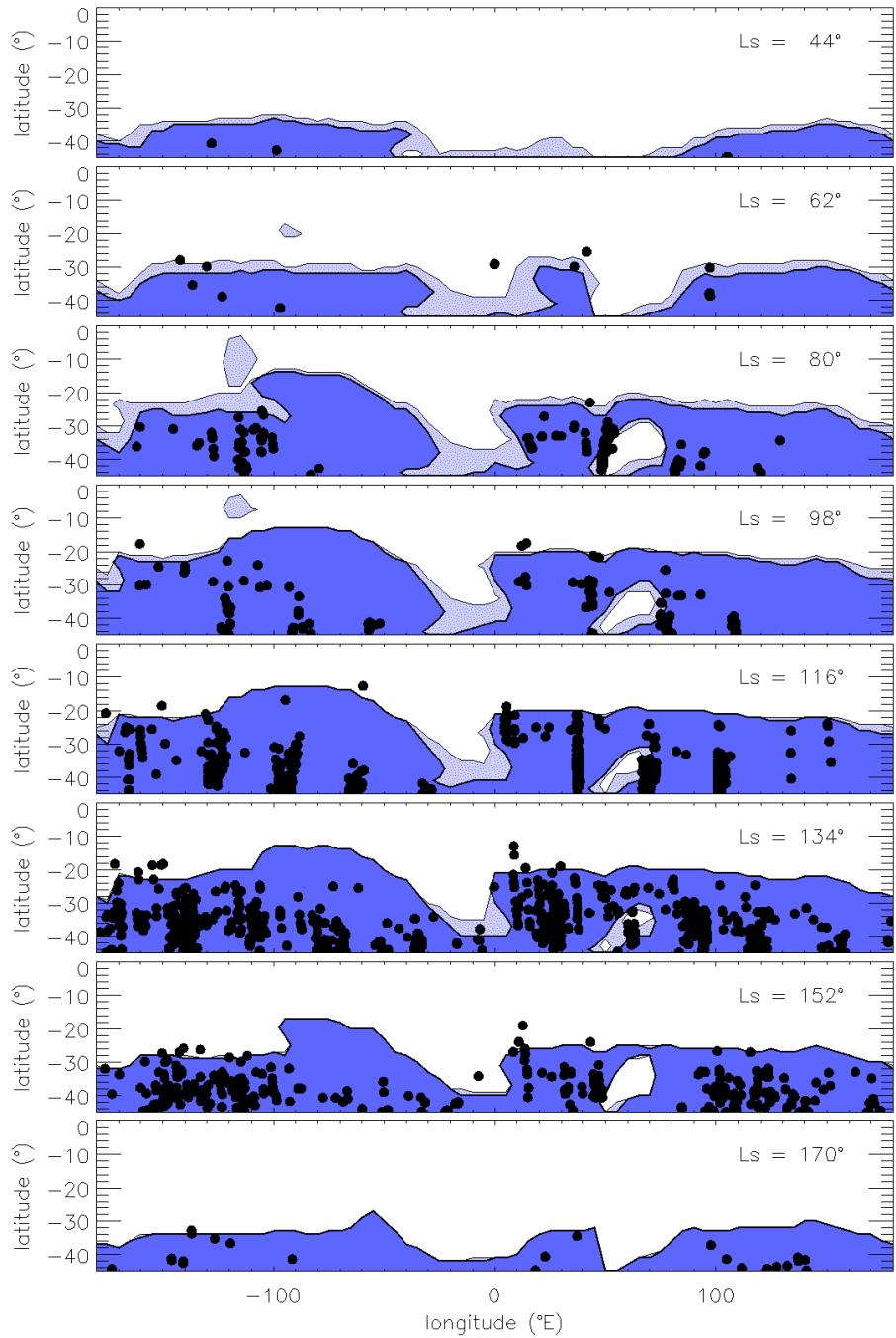

**Figure 5.** *Predicted area of ice deposits on slopes (dark blue, 5μm thick and more; light blue, 2μm thick) compared to observations (dots). Model predictions are shown for 8 solar longitudes. Corresponding observations at ± 9° of $L_S$ are indicated.*

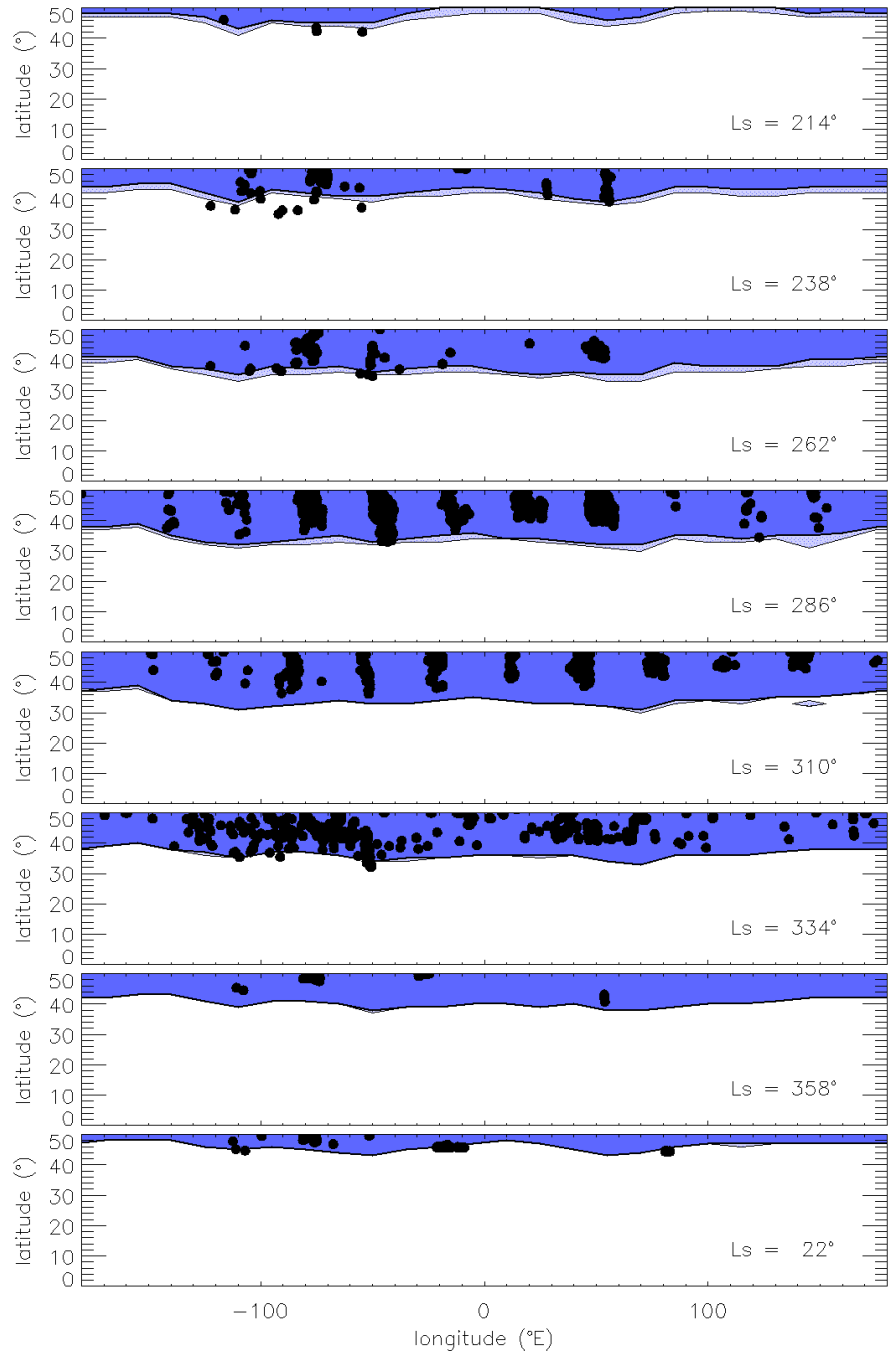

**Figure 6.** *Same as Figure 5, but for the Northern Hemisphere.*

### 4.1. Large-scale variability: Southern Hemisphere

In the southern hemisphere (Figure 5), the major longitudinal patterns are well reproduced by the model. The observed dry area between 50°W and 0° is notably predicted.

The origin of this asymmetry is illustrated in Figure 7, which shows the wind and atmospheric water content above the surface in early winter. Between 50°W and 0°, the atmospheric circulation is characterized by strong northward flux in the lower branch of the Hadley cell, concentrated in this longitudinal range by the phenomena of western boundary current on the eastern side of the Tharsis bulge [*Joshi et al.*, 1994; 1995]. At this season the amount of atmospheric water is much lower in the Southern Hemisphere (winter). Consequently, the northward flux is relatively dry and reduces the stability of surface water ice. Similarly, the Hellas topographic depression induces a stationary wave, which generates a northward flux of dry air into the Hellas basin. As a results, a lack of surface frost is predicted in north-west Hellas (50°E – 70°E / 30°S – 45°S) and is consistent with observations.

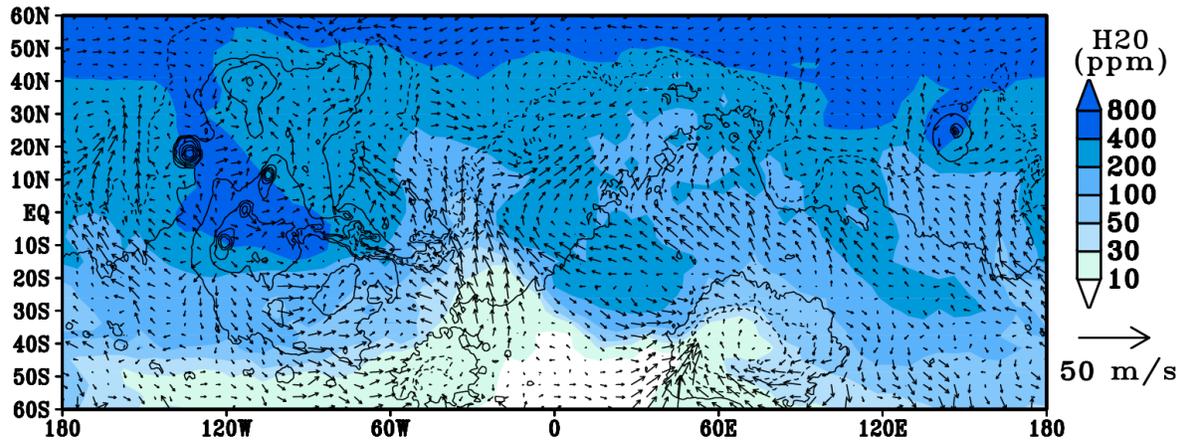

**Figure 7.** *Mean wind and atmospheric $H_2O$ content (vapor and ice) at 100 m above the surface, averaged over $L_S$ 90°–$L_S$ 120° (early southern winter).*

The model routinely predicts ice at low latitudes (down to 13°S) in the Thaumasia area (110°W–60°W) while only a few spots are observed there. Potential thin (~2 μm) ice deposits are also predicted near Arsia Mons, while not observed. Several factors contribute to this discrepancy. First, the region is covered by Hesperian lava and contains fewer slopes compared to the highly cratered Noachian terrains found at other longitudes, which reduce the probability to observe ice. Second, only the steepest slopes (30°) can hold water ice at most $L_S$ in this area (Figure 8). Finally, the model over-predicts cloud opacity and therefore precipitation from the aphelion cloud belt [*Montmessin et al.*, 2004]. This is illustrated in Figure 9 where the predicted thickness of ice deposits is mapped. While the density of positive water ice observations generally correlate with the predicted thickness (up to about 100 μm for most longitudes), very thick deposits (> 150 μm) are predicted in Thaumasia where only a few observations of ice are reported.

The local latitude maximum observed at 10°E (13°S, compared to about 20°S for surrounding longitudes) is not reproduced by the model with the default set of parameters described in section 3.2. While specific local conditions on slopes (inertia, albedo) favoring surface ice formation can easily explain ice at these lower latitudes (as discussed below), the "peak" structure in the observations (with an equatorward limit that progressively increases from 5°E to 10°E and then decrease from 10°E to 15°) suggests a possible local meteorological condition not reproduced by the model at its 5° spatial resolution.

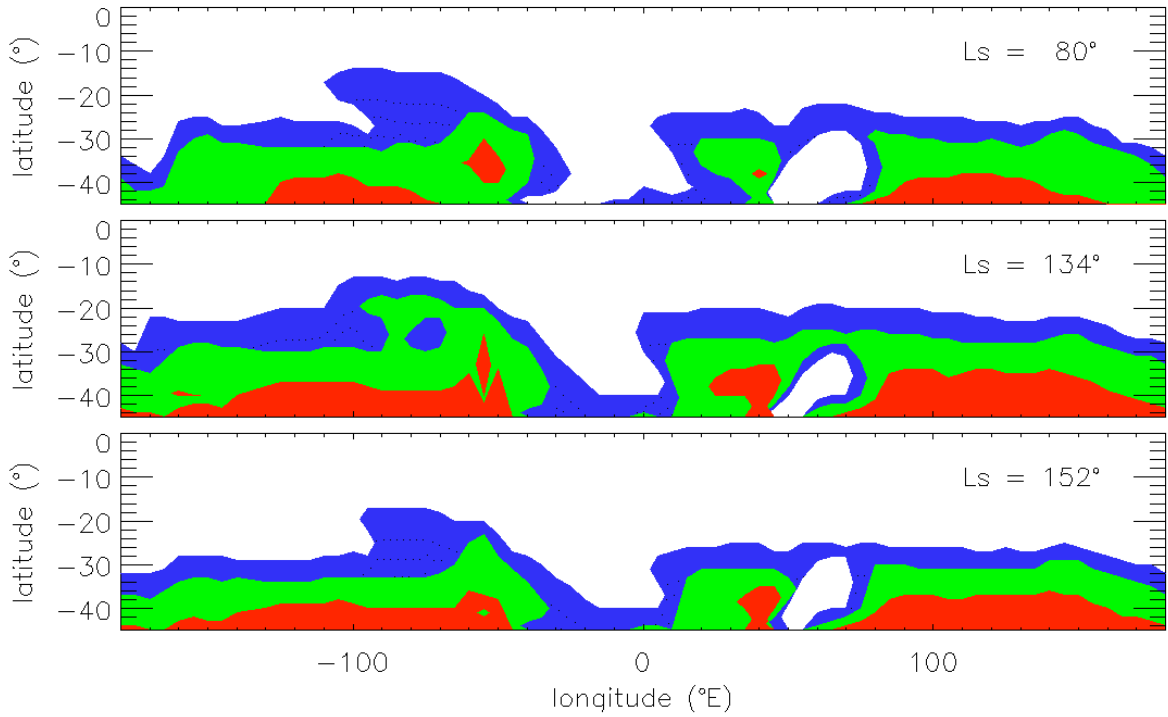

**Figure 8.** *Stability limit of 5 μm thick water ice deposits predicted by the model as a function of slope angle (blue, 30°, green, 20° and red, 10°) for 3 $L_S$.*

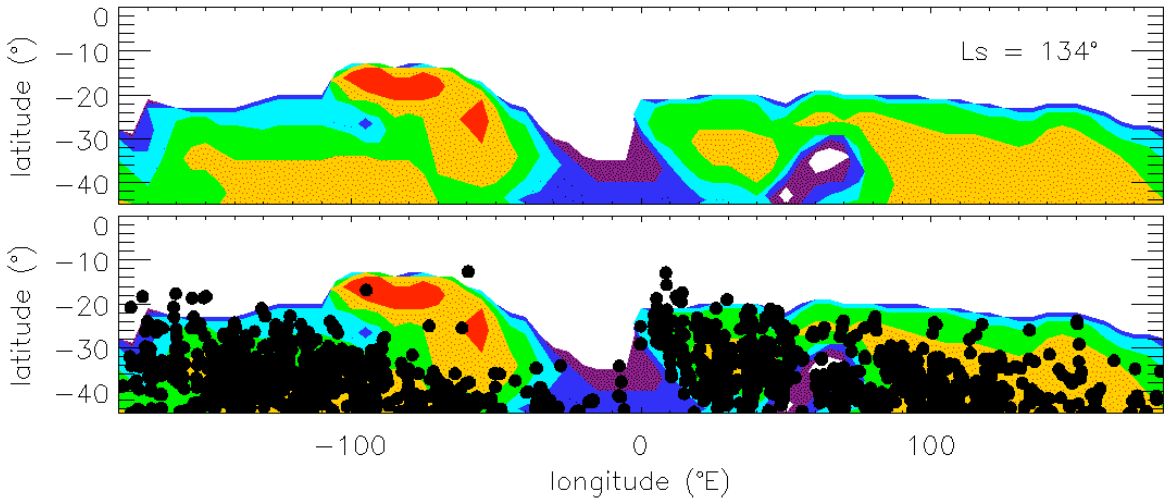

**Figure 9.** (top) *Thickness of water ice deposits predicted by the model at $L_S$ 134° in the Southern Hemisphere. Purple, 2 μm; dark blue, 5 μm; light blue, 10 μm; green, 20 μm; yellow, 50 μm; red, 150 μm. (bottom) Observed deposits (all $L_S$) are superimposed for comparison.*

## 4.2. Large-scale geographic variability: Northern Hemisphere

The Northern Hemisphere (Figure 6) contrasts with the Southern Hemisphere as it is characterized by smooth longitudinal variations of water ice stability both in the data and in the model. The equatorward limit of ice fluctuates mainly between 30°N and 40°N. The driest place is the smooth plain of Amazonis Planitia, with ice not observed equatorward of 45°N. While the presence of fewer favorable slopes in this area could explain the results, the model indeed predict a local stability minimum there. The effect of a lower density of slopes can be seen in Utopia planitia (80°E–150°E): observed deposits are seen at all latitudes where the model predicts ice, but detections are sparse. The Northern Hemisphere is characterized by a quite uniform thickness distribution: during fall the equatorward limit of "2μm" deposits is only 2° further the "5 μm" limit, and both thickness limits are rapidly mixed up as we enter winter. Deposits reach a maximum thickness of about 20–50 μm at most longitudes and latitudes after the northern winter solstice.

## 4.3. Ground properties below the ice: thermal inertia, albedo, slope angle

In the southern hemisphere, a few ice deposits are observed at latitudes significantly lower than predicted (Figure 5): about 17°S in the 150°W–170°W range, and 13°S at 10°E (while the limit predicted using the default set of parameters is about 19°S). The use of high surface inertia (1200 kg K$^{-1}$ s$^{-5/2}$), corresponding e.g. to boulder/rock [*Christensen et al.*, 2003], makes it possible to increase the equatorward limit of the stability of ice by 5° (Figure 10a), providing a possible explanation for those deposits. Increasing the albedo of the ground below the ice from 0.15 to 0.35 shifts the equatorward limit of ice stability by 4° (Figure 10b). A localized increase of surface albedo could thus also contribute to locally shift the equatorward limit of ice stability.

In the southern hemisphere, ice is observed up to L$_S$ 170°, in agreement with modeling predictions. The model predicts surface ice up to L$_S$ 220° if we remove subsurface ice (Figure 10a). Indeed, without subsurface ice the model over-predicts the condensation of CO$_2$ ice [*Vincendon et al.*, 2010], a cold trap for H$_2$O ice. While CO$_2$ observations were only sensitive to the upper limit of the depth of subsurface ice at 25°S (found to be about 90 cm, [*Vincendon et al.*, 2010]), H$_2$O ice provides a constraint on the lower limit of subsurface ice: a layer of ice too close to the surface (< 10 cm) indeed reduces the equatorward limit of surface water ice too far (Figure 10a), which is consistent with the expected deeper permafrost at such low latitudes [*Aharonson and Schorghofer*, 2006]. Model results are not so sensitive to subsurface ice in the Northern Hemisphere (Figure 11), in particular because CO$_2$ ice does not condense on slopes (the solar flux during the northern winter solstice – almost concomitant with perihelion – is about 50% higher than for the southern winter solstice). Similarly to the Southern Hemisphere, a very shallow (< 10 cm) widespread layer of subsurface ice below steep slopes is not consistent with observations at low latitudes (30°N–35°N).

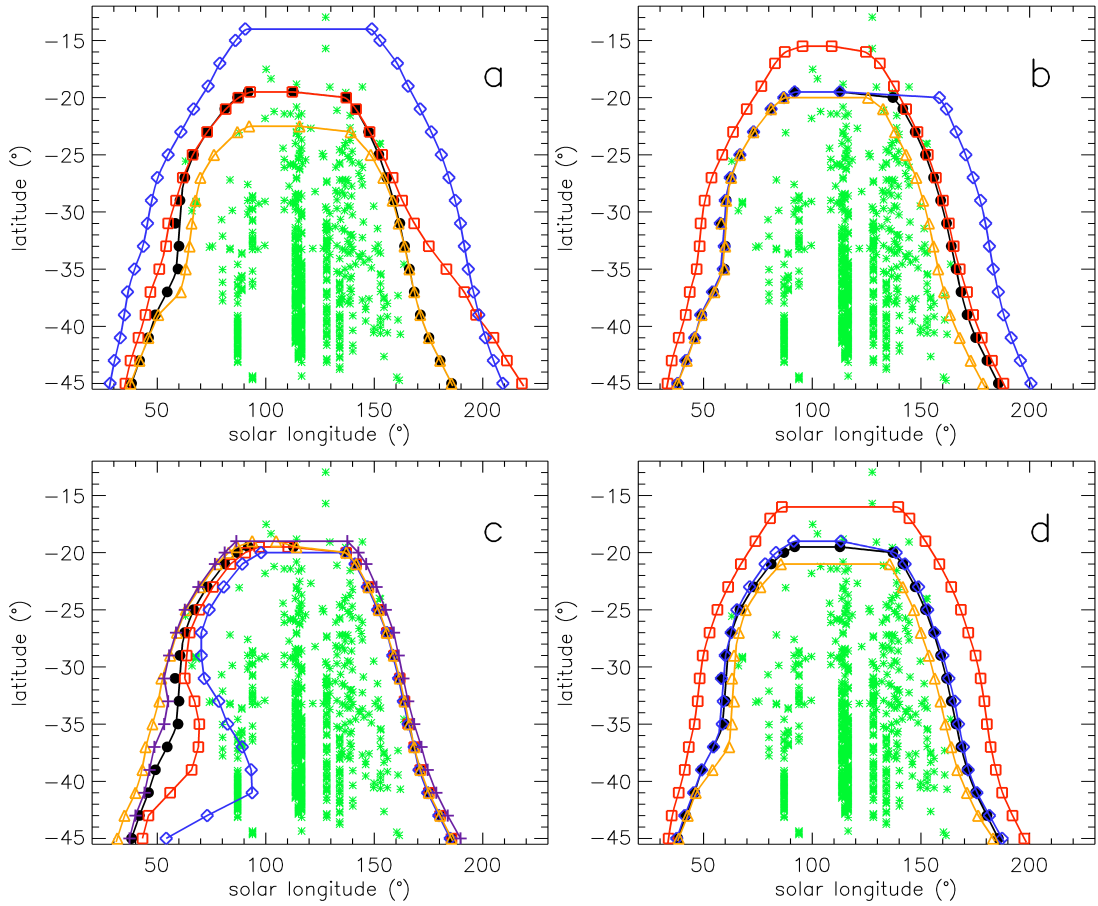

**Figure 10.** *Model compared to observations in latitude/season diagrams at 10°E in the Southern Hemisphere. Observed water ice deposits (green stars) correspond to the 0°–60°E longitude range. For all panels, black dots correspond to the standard parameters set as described in section 3.2. (**a**) Impact of surface/subsurface thermal inertia I (standard: I = 250 kg $K^{-1}$ $s^{-5/2}$ above I = 2120 kg $K^{-1}$ $s^{-5/2}$ – subsurface ice, with a depth varying from 6 cm at 45°S to 90 cm at 25°S). Red squares, no subsurface ice; orange triangles: subsurface ice at a constant depth of 10 cm; blue diamonds, I = 1200 kg $K^{-1}$ $s^{-5/2}$ for both surface and subsurface. (**b**) Impact of albedo (standard: substrate albedo $A_S$ of 0.15 and ice albedo $A_L$ of 0.4). Orange triangles, $A_L$ = 0.25; blue diamonds, $A_L$ = 0.60; red squares, $A_S$ = 0.35. (**c**) Impact of water ice thickness thresholds and dust aerosols (standard: threshold T = 2 µm and Opportunity dust scenario). Orange triangles, T = 1 µm; red squares, T = 3 µm; blue diamonds, T = 5 µm; purple crosses, Spirit dust scenario (optical depth about 30% lower in southern winter). (**d**) Impact of winds (standard wind speed: v = 20 m.$s^{-1}$). Blue diamonds, v = 10 m.$s^{-1}$; orange triangles, v = 50 m.$s^{-1}$; red squares, v = 1 m.$s^{-1}$.*

Insights into the dependence of model results to slope angle are provided in Figure 8 and 11. In the Northern Hemisphere (Figure 11), the equatorward limit of ice stability is reduced by about 3°–5° for every 10° decrease of the slope angle. In the Southern

Hemisphere, the decrease with slope angle is more pronounced (typically about 5°–10° latitude per 10° of slope angle), with significant longitudinal variations (Figure 8).

### 4.4. Ice properties

Changing the albedo of ice (Figure 10b) does not impact the starting date of the condensation, which depends on the properties of the slope before ice formation. On the contrary, increasing the albedo of ice significantly delays the sublimation (by about 20°–30° of $L_S$ between albedo of 0.25 and 0.60). Low to moderate ice albedos (0.3–0.4) are more consistent with observations (Figure 10b), in agreement with the expected properties of water ice: observed seasonal frost is generally low albedo (see section 3.2 and [*Vincendon et al.*, 2007]), and our model shows that most of the ice on slopes is formed via precipitation of atmospheric water ice, which is known to contain atmospheric dust [*Vincendon et al.*, 2008].

The thickness threshold plays a significant role in modeling predictions in fall in the Southern Hemisphere when condensation starts (Figure 10c). Once 1μm of water ice as accumulated at the surface, a 50° $L_S$ period is needed to reach 5μm at certain latitudes and longitudes, whereas the accumulation is significantly faster (10° $L_S$ only) at other locations. Comparisons of modeling predictions with observations (Figure 10c) show that the relevant threshold is between 2μm and 5μm, in agreement with expected instrumental capabilities (see section 3.2). The exact value of this threshold is a complex function of observation conditions (photometric angles, atmospheric conditions, spatial sampling, etc.). Two thresholds (2 and 5 μm) are therefore shown in maps of Figures 5 and 6 to assess the influence of this parameter as a function of location. The thickness of deposits is also mapped on Figure 8 for $L_S$ 134° (at this time most deposits have reached their maximum thickness). The thickness of most deposits on 30° pole facing slopes is 20–50 μm for both the Southern and the Northern Hemisphere.

### 4.5. Impact of aerosols and wind

A significant part of incoming solar radiations on pole-facing slopes is scattered by aerosols in winter at midlatitudes. This component of the radiative budget must therefore be modeled properly, as shown by modeling results corresponding to $CO_2$ ice [*Vincendon et al.*, 2010]. Uncertainties in the amount of atmospheric dust during winter are the most prominent source of model uncertainties related to aerosols [*Vincendon et al.*, 2010]. The optical depth is low in southern winter while northern winter corresponds to the storm season. Optical depth measurements by Spirit and Opportunity [*Lemmon et al.*, 2004] provide a good proxy for low to midlatitudes [*Vincendon et al.*, 2009]. We found that results are only changed by a few degrees of latitude or $L_S$ if we change assumptions regarding the amount of aerosols in the atmosphere (Figure 10c and Figure 11).

Wind speed controls the sublimation flux of surface ice when the atmosphere is not saturated in water vapor. A constant wind speed is used in our modeling approach, with a default value set to 20 m.s$^{-1}$ (see section 3.2). As expected, lower winds increase the stability of ice (Figure 10d). Model results are however weakly modified by a change in the wind speed assumption: results are similar for winds in the 10-50 m.s$^{-1}$ range expected from mesoscale modeling [*Spiga et al.*, 2010]. Very low local winds (~ 1 m.s$^{-1}$) increases the stability of ice by about 3° latitude and 10–20° of L$_S$, which could also contribute to explain local mismatch of that order between modeling predictions and observations (see section 4.3).

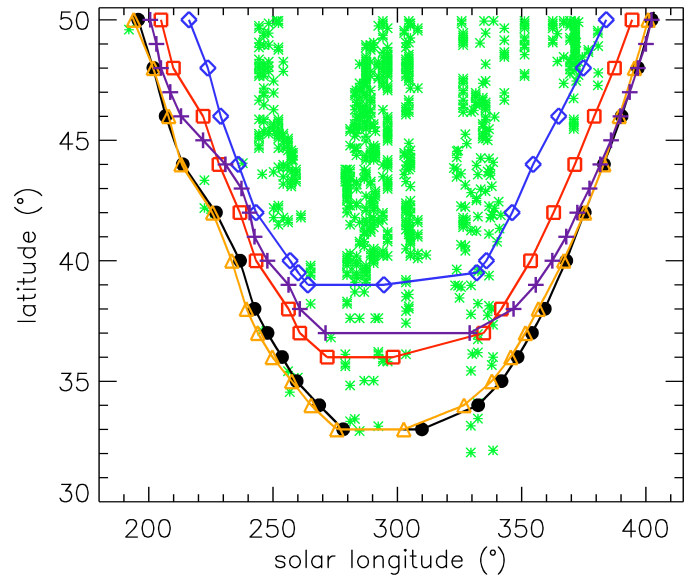

**Figure 11.** *Model compared to observations in latitude/season diagrams at 50°W in the Northern Hemisphere with variations in dust aerosols, slopes angle, and subsurface ice. Observed water ice deposits (green stars) correspond to the 80°W–20°W longitude range. Black dots, standard scenario (30° slope angle, Opportunity MY27 dust scenario, no subsurface ice). Orange triangles, Opportunity MY28 dust scenario, including a global dust storm in northern winter. Purple crosses, subsurface water ice at a constant depth of 10 cm is added. Red squares, 20° slope angle. Blue diamonds, 10° slope angle.*

**Conclusion**

We used near-IR observations by CRISM and OMEGA to map as a function of time the presence of surface water ice during the day at low and midlatitudes (45°S–50°N latitude range). These instruments have observed Mars for 2 and 3 Mars years, respectively, at all locations and seasons, providing a huge data set exempt of major observational bias. Deposits as thin as 2 µm are detected using the vibrational absorptions of water ice located at a wavelength of 1.5 µm. Surface ice is confidently identified among potentially cloudy observations through the use of spatial extent considerations. At these latitudes, ice is observed on pole-facing slopes. Most deposits are found on crater rims; however ice is observed on any kind of slope such as those created by escarpments, mesas and valleys. In the Northern Hemisphere, the equatorward limit of ice shows smooth variations with longitude, with a local maximum of 32°N observed at 50°W and a local minimum near 45°N in Amazonis. The Southern Hemisphere is characterized by a much more complex geographic pattern: ice extends at significantly equatorward latitudes (13°S at 10°E and in Valles Marineris) but is only observed down to 40°S at 20°W. 13°S is the equatorward limit of surface ice reported so far.

We developed a modeling approach based on the LMD GCM to predict the stability of water ice and analyze our observations. Our model combines a 1-D energy balance code that can be used to compute the temperature on slopes with seasonal maps of water vapor and water ice precipitation predicted by the 3-D GCM. Overall, we found a very good agreement between modeling predictions and observations. The model shows that the large-scaled geographic variability in surface water ice is mainly driven by the global meteorology. A particularly dry area is notably found in the Southern Hemisphere between longitudes 50°W and 0° and explained by flux of dry air during winter over this area. While being of second order, local properties also impact ice condensation, which is favored on high inertia and high albedo surfaces. A low albedo for water ice (0.4 or less) is most consistent with observations according to the model. Both observations and modeling predictions shows that local time during daylight does not significantly modify the distribution of detectable ice deposits: the ice thickness required for ice to be detected (2–5 µm) is such that ice needs to accumulate several days to be seen. According to the model, deposits are typically a few tens of µm thick.

Surface ice has only been found in fall and winter (as well as in early spring for the Northern Hemisphere) at times and places totally consistent with seasonal ice formation predicted by the model. No exposed perennial ice has hence been detected, which is expected at those latitudes. The presence of perennial ice in the shallow subsurface strongly increases its thermal inertia, which significantly impacts surface temperatures. Surface temperatures then modify the capability of gaseous species ($CO_2$ and $H_2O$) to condense [*Kossacki and Markiewicz*, 2002]. Observations of surface $CO_2$ ice combined with GCM predictions have been previously used to detect the presence of subsurface water ice down to 25° latitude in the southern hemisphere [*Vincendon et al.*, 2010]. Our present study provided interesting complementary evidence for such subsurface ice. First, the same model and parameters used to match $CO_2$ ice observations are successfully used here to model $H_2O$ ice without adjustments. In particular, the study of $H_2O$ ice has provided an opportunity to verify the ability of the model to match frost observations where subsurface ice can not be invoked (i.e., at latitudes as low as 13°). Second, observations of $H_2O$ ice are also, to a lower extent, sensitive to subsurface ice. They notably provides a lower limit for the ice table depth at 25°S (10 cm), which completes the upper limit (90 cm) estimated using $CO_2$ frost. In the Northern Hemisphere, subsurface ice on steep pole facing slopes, if present, have to be buried deeper than 10 cm at 30°–35° latitude.

**Acknowledgment:**
We would like to thank the LMD GCM, the OMEGA and the CRISM engineering and scientific teams, for their help and for making this work possible. We thank in particular (in alphabetical order) Francesca Altieri, Jean-Pierre Bibring, Giacomo Carrozzo, Brigitte Gondet, Misha Kreslavsky, Yves Langevin, Jean-Baptiste Madeleine, Ehouarn Millour, and Aymeric Spiga.